\documentclass[prb,twocolumn,showpacs,aps]{revtex4}

\usepackage{graphicx}%
\usepackage{dcolumn}
\usepackage{amsmath}

\begin{document}

\preprint{Resubmitted to Phys. Rev. B - Brief Reports}

\title{Viscoelastic response of sonic band-gap materials}

\author{I.E.~Psarobas}

\thanks{Also at the Section of Solid State Physics of the University of Athens,
Panepistimioupolis, GR-157 84, Athens, Greece.}

\email{ipsarob@cc.uoa.gr} \homepage{http://www.uoa.gr/~vyannop}

\affiliation{Department of Physics, National Technical University
of Athens\\ Zografou Campus, GR-157 73, Athens, Greece}
\date{\today}
\begin{abstract}
A brief report is presented on the effect of viscoelastic losses
in a high density contrast sonic band-gap material of
close-packed rubber spheres in air. The scattering properties of
such a material are computed with an on-shell multiple scattering
method, properties which are compared with the lossless case. The
existence of an appreciable omnidirectional gap in the
transmission spectrum, when losses are present, is also reported.
\end{abstract}


\maketitle
The problem of elastic wave propagation in inhomogeneous media is
of great importance in many branches of physics, mathematics and
engineering. In particular, matters such as the localization of
classical waves~\cite{PSheng} and the formation of spectral gaps
in periodic elastic composites,~\cite{Sig1,Kushw} have drawn the
attention of researchers over the last decade. In particular,
phononic or sonic crystals are composite materials which consist
of homogeneous particles (solid or fluid inclusions the
dimensions of which are large enough in order for a macroscopic
description of their elastic properties to be valid) distributed
periodically in a host medium characterized by different mass
density and Lam\'e coefficients. With an appropriate choice of
the parameters involved one may obtain sonic crystals with
absolute frequency gaps (omnidirectional sonic gaps).

Among the various methods available for the calculation of the
elastic properties of phononic crystals, the traditional
band-structure methods mainly deal with periodic, infinite, and
nondissipative structures. However in an experiment, one deals
with finite-size slabs and the measured quantities are, usually,
the transmission and reflection coefficients. Apart from that,
realistic structures are dispersive and have losses. This
limitation of traditional band-structure calculations was noticed
in a theoretical study of colloidal crystals with
ultrasound.~\cite{Sprik} We remember that a usual band-structure
calculation proceeds with a given wave vector, ${\bf k}$, and
computes the eigenfrequencies within a wide frequency range
together with the corresponding eigenmodes. On the contrary,
on-shell methods proceed differently: the frequency is fixed and
one obtains the eigenmodes of the crystal for this frequency.
These methods are ideal when one deals with dispersive materials
(with or without losses). Moreover on-shell methods are
computationally more efficient than traditional band-structure
methods.~\cite{Leung} Psarobas {\em et al.} have recently
developed an on-shell method for phononic
crystals,~\cite{ipsarob1} which applies to systems which consist
of nonoverlapping homogeneous spherical particles arranged
periodically in a host medium characterized by different elastic
coefficients. The method provides the complex band structure of
the infinite crystal associated with a given crystallographic
plane; and also the transmission, reflection, and absorption
coefficients of an elastic wave incident at any angle on a slab
of the crystal, parallel to a given plane, of finite thickness.

The present work introduces the effect of viscoelasticity in a
sonic band-gap material. For this purpose, a binary system of
close-packed rubber spheres in air is chosen. The viscoelastic
response of the system is accounted for by means of the
Kelvin-Voigt model,~\cite{Gaunard} which is well-suited for
materials and ultrasonic frequencies of major interest. The
problem of acoustic-wave scattering by a single viscoelastic
sphere of radius $S$ has been adequately addressed in the
past~\cite{Gaunard} according to the Kelvin-Voigt viscoelastic
model. In such a case the sphere is considered to be elastic with
modified shear and compressional complex wavenumbers, the
imaginary parts of which represent a measure of the loss. In
particular, for an absorbing sphere in an inviscid fluid
background, the complex compressional and shear wavenumbers are
conveniently defined as
\begin{eqnarray}
q_{sl}&=& \frac{c_l}{c_{sl}}\ \frac{q_l}{\sqrt{1-i[(\alpha +
\beta)/\rho_s c_{sl}^2]}}\;, \nonumber \\
q_{st}&=& \frac{c_l}{c_{st}}\ \frac{q_l}{\sqrt{1-i(\beta /\rho_s
c_{st}^2)}}\;, \label{eq:viscq}
\end{eqnarray}
where $q_l=\omega /c_l$ refers to the fluid environment with
$\omega$ being the angular frequency and $c_l$ the respective
speed of sound. The real parts of the complex Lam\'e parameters of
the sphere, $\lambda_s=\lambda_{se}-i\lambda_{sv}$ and
$\mu_s=\mu_{se}-i\mu_{sv}$, combined with the density $\rho_s$
yield the compressional and shear wave speeds respectively
\begin{equation}
c_{sl}=\sqrt{(\lambda_{se}+2\mu_{se})/\rho_s}\,,\ \
c_{st}=\sqrt{\mu_{se}/\rho_s}\,. \label{eq:wavespeeds}
\end{equation}
The imaginary parts of the Lam\'e parameters are connected to the
viscous losses $\alpha+2\beta$, and $\beta$ of the sphere as
follows: $\alpha=\omega \lambda_{sv}$, $\beta=\omega \mu_{sv}$.
The problem of elastic scattering by a solid sphere in an inviscid
fluid~\cite{Brill} is described by the scattering transition
matrix, the elements of which, in the angular momentum
representation $(l,m)$ (see Appendix), connect the spherical wave
expansion coefficients~\cite{ipsarob1} of the scattered field to
those of the incident.

Multiple scattering effects within planes of spheres, sonic
crystals and slabs of the same are taken into account by the
method described in Ref.~\onlinecite{ipsarob1}. This method views
the crystal as a sequence of planes of spheres parallel to a given
surface: a crystallographic plane described by a two-dimensional
(2D) lattice $\{{\bf R}_n\}$. The corresponding 2D reciprocal
lattice we denote by $\{{\bf g}\}$. In the host region between the
$n$-th and the $(n+1)$-th planes, a Bloch wave solution for the
displacement field (harmonic time dependence is assumed),
corresponding to a given frequency $\omega$ and a given reduced
wave vector ${\bf k}_{\|}$ within the surface Brillouin zone
(SBZ) of the given surface, can be expanded into plane waves
propagating (or decaying) to the left and to the right, as
follows:
\begin{eqnarray}
{\bf u}(\omega;{\bf k}_{\|})= & & \sum_{{\mathbf{g}}} \bigl\{{\bf
u}_{{\mathbf{g}}n}^{+}
\exp\left[i\,{\mathbf{K}}_{{\mathbf{g}}l}^{+} \cdot
({\mathbf{r}}-{\mathbf{A}}_n)\right]\nonumber \\ & &+\ {\bf
u}_{{\mathbf{g}}n }^{-}
\exp\left[i\,{\mathbf{K}}_{{\mathbf{g}}l}^{-} \cdot
({\mathbf{r}}-{\mathbf{A}}_n)\right]\bigr\}\;, \label{eq:unn+1}
\end{eqnarray}
where
\begin{equation}
{\bf K}_{{\bf g}l}^{\pm}= \left({\bf k}_{\|}+{\bf g},\,\pm
\left[(\omega /c_l)^2-({\bf k}_{\|}+{\bf
g})^2\right]^{1/2}\right) \label{eq:Kg}
\end{equation}
and ${\bf A}_n$ is a point between the $n$th and $(n+1)$th
planes. We note that the separation of two successive planes of
spheres need only be larger than the radii of the spheres of the
two planes (see pg. 86 of Ref.~\onlinecite{Modinos}). It should
be also noted that, although both shear and compressional modes
are considered within the spheres and enter the calculation
through the scattering transition matrix (see Appendix), only
longitudinal waves exist in the host (air) region [see
Eq.~(\ref{eq:unn+1})]. As a consequence of that, in a binary
composite of nonoverlapping solid spheres in a fluid host, such as
the one being investigated in this report, there are no
propagating shear waves if the solid component does not form a
continuous network.

A generalized, i.e. propagating or evanescent, Bloch wave
satisfies the equation
\begin{equation}
{\bf u}_{{\mathbf{g}}n+1}^{\pm}= \exp\left(i{\mathbf{k}}
\cdot{\mathbf{a}}_{3}\right) {\bf u}_{{\mathbf{g}}n}^{\pm}\,,
\label{eq:blochwave}
\end{equation}
where ${\bf a}_3={\bf A}_{n+1}-{\bf A}_n$ and
${\mathbf{k}}=\left({\mathbf{k}}_{\|},k_{z}(\omega;{\mathbf{k}}_{\|})
\right)$ is the Bloch wavevector. There are infinitely many such
solutions for given ${\bf k}_{\|}$ and $\omega$, corresponding to
different values of the $z$-component, $k_z(\omega;{\bf
k}_{\|})$, of the reduced wave vector ${\bf k}$, but in practice
one needs to calculate only a finite number (a few tens at most)
of these generalized Bloch waves. We have propagating waves [for
these $k_z(\omega;{\bf k}_{\|})$ is real] which constitute the
normal modes of the infinite crystal; and evanescent waves [for
these $k_z(\omega;{\bf k}_{\|})$ is imaginary] which do not
represent real waves, but they are useful mathematical entities
which enter into the evaluation of the reflection and transmission
coefficients of a wave, with the same $\omega$ and ${\bf
k}_{\|}$, incident on a slab of the crystal parallel to the given
surface. The transmission/reflection matrices for a slab which
consists of a stack of layers of spheres with the same 2D
periodicity parallel to a given plane of the crystal are obtained
from the transmission/reflection matrices of the individual
layers in the manner described in Ref.~\onlinecite{ipsarob1}.
Knowing the transmission/reflection matrices for the slab we can
readily obtain the transmission, reflection, and absorption
coefficients of a plane acoustic wave incident on the slab.

The system which will be examined here is a crystal of rubber
spheres in air. The physical parameters entering our calculations
are taken from Ref.~\onlinecite{Gaunard}. In particular, the mass
density of air is $\rho_{air}=1.2\ {\rm kg/m^3}$  and its
respective speed of sound $c_{air}=334\ {\rm m/s}$. The rubber
spheres have a mass density $\rho_s=1130\ {\rm kg/m^3}$ and
$c_{ls}=1400\ {\rm m/s}$, $c_{ts}=94\ {\rm m/s}$ are the
compressional and shear speeds of sound, respectively. In
addition, according to Ref.~\onlinecite{Gaunard}, three different
viscoelastic levels are considered for the rubber spheres,
namely: lossless spheres ($\alpha=\beta=0$), a low viscous level
($\alpha_{low}=0.5\ {\rm MPa/s}$, $\beta_{low}=0.01\ {\rm
MPa/s}$) and a high viscous level ($\alpha_{high}=5\ {\rm
MPa/s}$, $\beta_{high}=0.1\ {\rm MPa/s}$). The viscoelastic
properties used in this study are typical values for commercial
rubbers, the variety of which is quite extensive~\cite{matweb} and
frequency dependent at high ultrasonic frequencies.~\cite{ucom}
\begin{figure}
\centerline{\includegraphics*[height=6cm]{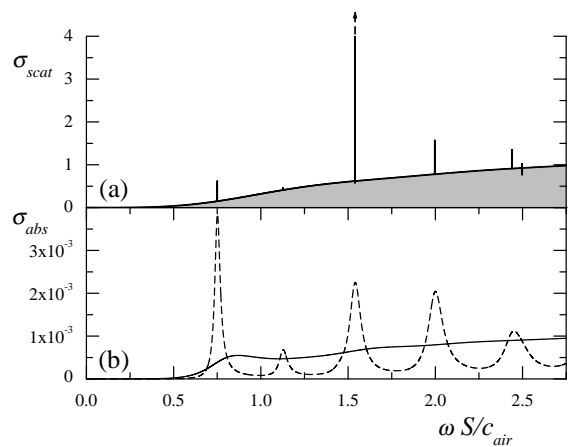}} \caption{(a):
Normalized total scattering cross section of a rubber sphere in
air. The solid line corresponds to a sphere with no losses and
the shaded curve to a sphere with losses. The two different
viscous levels used here, for the sphere, are hardly
distinguishable. The resonance with the arrow extends twice as
high. (b): Normalized absorption cross section of a viscoelastic
rubber sphere. The dotted (solid) line corresponds to the low
(high) viscous level used here.} \label{fig1}
\end{figure}

Before dealing with the main object of this work, it will be
useful to state briefly certain basic features of the acoustic
scattering problem by a viscoelastic sphere.~\cite{Gaunard} These
features will be used extensively in what follows. These are
presented in Fig.~\ref{fig1}, obtained with an angular momentum
cutoff, $l_{max}=5$, which yields an accuracy of better than
0.001\% in the given frequency range. On first sight, one may
observe the disappearance of the sharp modal resonances in the
scattering cross section of the sphere (see Appendix), when
viscoelasticity is turned on. Also, there is no significant
difference between the two viscous levels (low and high) in
scattering. On the contrary absorption (see Appendix) seems to be
different for the two cases. The low viscous level induces
resonant absorption exactly at the resonant frequencies of the
lossless case, while the higher viscous level washes everything
out as if there was no inner resonant structure in the system.

\begin{figure}
\centerline{\includegraphics*[height=6cm]{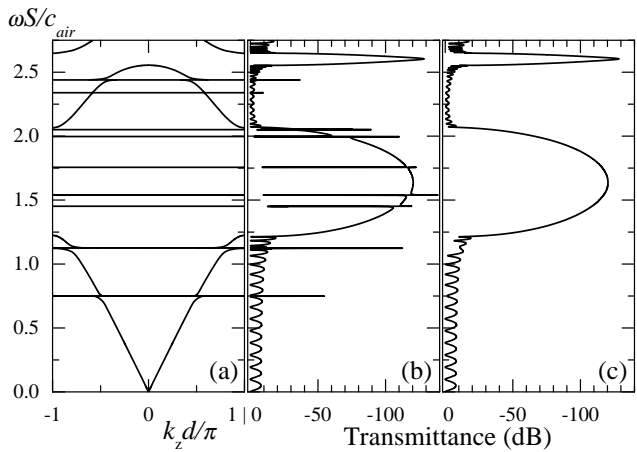}} \caption{The
sonic band structure at the center of the SBZ of a (111) surface
of an fcc crystal of close-packed lossless rubber spheres in air
(a). The corresponding transmittance curve of a slab of 16 layers
parallel to the same surface is given in (b). In (c) the same
transmittance curve is presented but with spheres of the low
viscous level. $d$ is the distance between successive (111)
planes of the fcc crystal under consideration.} \label{fig2}
\end{figure}

\begin{figure}
\centerline{\includegraphics*[height=6cm]{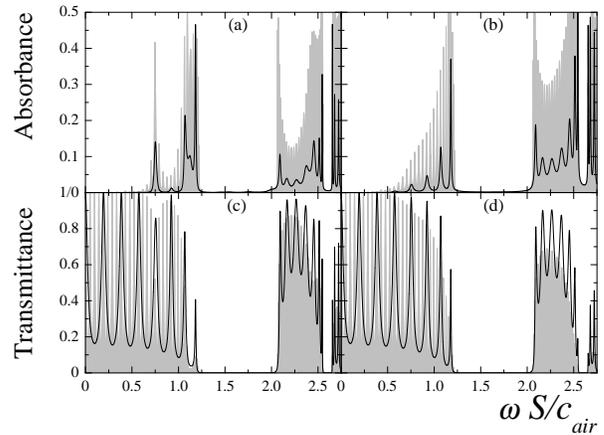}}
\caption{Absorbance and transmittance curves of slabs of the
rubber sonic crystal described in Fig.~\ref{fig2}(a) consisting
of 8 [(a),(c)] and 32 [(b),(d)] planes of spheres, respectively.
The black line (shaded curve) corresponds to the low (high)
viscous level.} \label{fig3}
\end{figure}

We next consider an fcc crystal of close-packed (almost touching)
rubber spheres in air. We view the crystal as a succession of
planes of spheres parallel to the (111) fcc surface.
Fig.~\ref{fig2}(a) shows the frequency band structure normal to
the fcc (111) plane (${\bf k}_{\|}={\bf 0}$) and the corresponding
transmission spectrum for waves incident normally on a slab of
the crystal consisting of 16 layers of lossless spheres. The
results are obtained with an angular momentum cutoff $l_{max}=7$
and 55 ${\bf g}$ vectors (the established convergence is within
an accuracy of better than a tenth of a percent). One observes,
besides a large Bragg gap extending from $\omega S/c_{air}=1.223$
to $\omega S/c_{air}=2.065$, a number of flat bands which derive
from the interacting sharp resonant modes localized on the
individual rubber spheres (see Fig.~\ref{fig1}).
Because these bands are so narrow in the present case, they are
hardly observable; except that they introduce small gaps, above
and below the main gap, which result from the hybridization of
these flat bands with the broad bands corresponding to nearly free
propagating waves. These narrow gaps are seen more clearly in the
transmission spectrum [Fig.~\ref{fig2}(b)]. Within the main gap
these flat bands manifest themselves as sharp peaks in the
transmission spectrum. The long wavelength limit is represented
by the linear segments of the dispersion curves of
Fig.~\ref{fig2}(a), the slopes of which determine the propagation
velocity of acoustic waves ($\overline{c}_l=1.54\;c_{air}$) in a
corresponding effective medium. The oscillations in the
transmission coefficient, over the allowed regions of frequency,
are due to interference effects resulting from multiple
reflection at the surfaces of the slab of the crystal
(Fabry-P\'erot-type oscillations). When losses are present in the
system, there are no true propagating waves and the band
structure of the infinite lossless crystal is not of any help,
therefore the effect of the low viscous level is shown in the
transmission spectrum [Fig.~\ref{fig2}(c)]. As expected from the
results of the single sphere, the sharp peaks and dips of the
resonant states disappear and we obtain a ``clean'' sonic gap
without any resonant modes within it. The existence of the
frequency gap means that sound does not propagate through the
crystal when its frequency lies within the gap (the intensity of
the wave decays exponentially into the crystal for these
frequencies), and if it cannot enter into the crystal, it cannot
be absorbed either. This is shown in Fig.~\ref{fig3}.

The close relation between absorption and transmission is
demonstrated in Fig.~\ref{fig3} for slabs consisting of 8 and 32
(111) planes of spheres, and for normal incidence. The solid lines
correspond to the lower viscous level and the shaded curves to
the higher one. A relatively large transmission coefficient
implies that a correspondingly large fraction of sound has gone
through the slab, with a consequent higher probability of being
absorbed. However, since losses are due to the rubber spheres,
absorption mainly occurs about the frequency regions where the
modes of the acoustic field are mostly localized in the spheres.
This explains why, outside the gap regions, absorption takes
place essentially about the frequencies of the flat bands.
\begin{figure}
 \centerline{\includegraphics*[height=6cm]{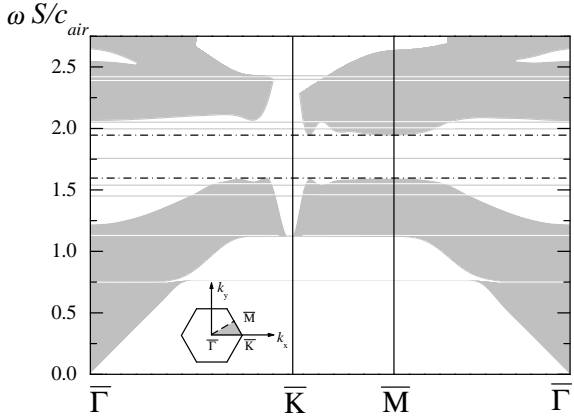}}
\caption{Projection of the frequency band structure on the SBZ of
the (111) surface of the fcc sonic crystal described in the
caption of Fig.~\ref{fig2}. The white areas show the frequency
gaps in the considered frequency region. The broken lines map the
omnidirectional frequency gap in the transmission spectrum of a
slab of the crystal, with losses, of finite thickness. Finally,
the inset shows the SBZ of the fcc (111) surface.} \label{fig4}
\end{figure}

In Fig.~\ref{fig4}, the projection of the frequency band structure
on the SBZ of the (111) plane of the fcc crystal along its
symmetry lines is shown. This is obtained, for a given ${\bf
k}_{\|}$, as follows: the regions of frequency over which there
are no propagating states in the infinite crystal [the
corresponding values of of all $k_z(\omega,{\bf k}_{\|})$ are
complex] are shown in white, against the shaded areas which
correspond to regions over which propagating states do exist [for
a given $\omega$ there is at least one solution corresponding to
$k_z(\omega,{\bf k}_{\|})$ real]. One clearly sees here how the
resonances on spheres lead to narrow hybridization gaps above and
below the main gap, and flat bands in the gap regions. When losses
are present, the crystal under investigation exhibits an
appreciable omnidirectional sonic transmission gap extending from
$\omega S/c_{air}=1.595$ up to 1.946. Finally, we note that the
results obtained depend on the ratio $\lambda_0/S$ ($\lambda_0$ is
the wavelength of the incident sound wave) and are therefore
applicable over different ranges of frequencies, provided that
the viscoelastic properties of the spheres do not vary
significantly with the frequency.

\begin{acknowledgements}
 This work has been supported by the Institute of Communication
 and Computer Systems (ICCS) of the National Technical University
 of Athens. Support from the University of Athens is also
 acknowledged.
 \end{acknowledgements}

\appendix
\section{}
\label{apxa} Applying the proper boundary conditions at the
interface between the surrounding fluid and the
sphere,~\cite{Brill} requiring the continuity of the radial
component of the displacement field and the surface traction,
along with the requirement that there is no tangential component
of the surface traction at the interface, we can determine the
scattering matrix which connects the incident with the scattered
field.  Here for reasons of completeness, along the lines of the
formalism established in Ref.~\onlinecite{ipsarob1}, the nonzero
elements of the {\bf T} matrix for a solid sphere in a fluid host
are
\begin{equation}
T_{l m; l'm'}^{LL}= \frac{W_{l}^{LL}}{D_l}\delta_{l l'}
\delta_{mm'}\;,\; l,l'\geq 0\;, \label{eq:Tmatrix}
\end{equation}
with $z_l=Sq_l$ referring to the fluid and $x_{\nu}=Sq_{s\nu}$,
with $\nu=l,t$ to the sphere. The superscripts $LL$ refer to the
case of scattering in a fluid host, since incident and scattered
waves are $L$-type compressional waves. The $3\times3$
determinants $D_{l}$ and $W_l^{LL}$ are given by
\begin{eqnarray}
D_{l}= \left|
\begin{array}{lll}
d_{22} & d_{23} & d_{24}\\ d_{32} & d_{33} & d_{34}\\
d_{42} & d_{43} & d_{44}
\end{array}
\right|\,,\ \  W_{l}^{LL}= \;\;\,-\left|
\begin{array}{lll}
d_{2}^L & d_{23} & d_{24}\\ d_{3}^L & d_{33} & d_{34}\\ d_{4}^L &
d_{43} & d_{44}
\end{array}
\right|\,, \label{eq:W}
\end{eqnarray}
where
\begin{equation}
\begin{array}{l}
d_{22}=z_l{h_{l}^{+}}'(z_l), \ \ d_{23}=l(l+1)j_{l}(x_t), \ \
d_{24}=x_lj'_{l}(x_l), \\  d_{32}=0, \ \ d_{33}=
\left[l(l+1)-1-x_{t}^2/2\right]j_{l}(x_t)-x_t j'_{l}(x_t), \\
d_{34}=x_lj'_{l}(x_l)-j_{l}(x_l), \ \ d_{42}=-x_t^2 \rho
h_{l}^+(z_l)/(2 \rho_s), \\  d_{43}=
l(l+1)\left[x_{t}j'_{l}(x_t)-j_{l}(x_t)\right], \\
d_{44}=\left[l(l+1)-x_t^2/2\right]j_{l}(x_l)-2x_lj'_{l}(x_l), \\
d_{2}^L=z_lj'_{l}(z_l), \ \  d_{3}^L=0, \ \  d_{3}^L=-x_t^2 \rho
j_{l}(z_l)/(2 \rho_s)\,.
\end{array}
\label{eq:dij}
\end{equation}
$j'_{l}$ and ${h_{l}^{+}}'$ denote the first derivatives of the
spherical Bessel and Hankel functions, respectively. The ${\bf
T}$ matrix, because of spherical symmetry is diagonal in $l$ and
independent of $m$.

The exact form of the above ${\bf T}$ matrix allows us to compute
the normalized total scattering cross section of an elastic
sphere (scattering cross section over $\pi S^2$) in a fluid host,
\begin{equation}
\sigma_{total}=\frac{4}{z_l^2}\sum_{l=0} (2l+1)\ \left|T_{l m;l
m}^{LL}\right|^2\,. \label{eq:scatcs}
\end{equation}
In addition the normalized absorption cross section is given by
\begin{equation}
\sigma_{abs}=-\frac{4}{z_l^2}\sum_{l=0} (2l+1)\ \left\{\left|T_{l
m;l m}^{LL}\right|^2+{\rm Re}[T_{l m;l m}^{LL}]\right\}\,.
\label{eq:abscs}
\end{equation}

{}

\end{document}